\documentclass{iaus}
\usepackage{graphicx}

\title[$^4$He in Extragalactic H~{\sc{ii}} Regions]
{Measurements of $^4$He in Metal-Poor Extragalactic H~{\sc{ii}} Regions: the Primordial Helium Abundance and the 
$\Delta Y/\Delta O$ Ratio}

\author[Peimbert {\it et al.}]
{M. Peimbert$^1$, A. Peimbert$^1$, L. Carigi$^1$, and V. Luridiana$^2$}
 
\affiliation{$^1$Instituto de Astronom\'{\i}a,
  Universidad Nacional Aut\'onoma de M\'exico,
  Apdo. postal 70-264, M\'exico D.F. 04510, Mexico\\
  email: {\tt peimbert@astroscu.unam.mx}\\
[\affilskip]$^2$Instituto de Astrof\'{\i}sica de Canarias, c/ V\'{\i}a L\'actea s/n, 38205 La Laguna, Spain}

\pubyear{2010}
\volume{268}
\pagerange{1-8}
\date{Nov, 2009}
\jname{Light elements in the Universe}
\editors{C. Charbonnel, M. Tosi, F. Primas, \& C. Chiappini, eds.}
\begin{document}

\maketitle

\begin{abstract}

We present a review on the determination of the primordial helium abundance
$Y_p$, based on the study of hydrogen and helium recombination lines in
extragalactic H~{\sc{ii}} regions. We also discuss the observational 
determinations of the increase of helium to the increase of oxygen
by mass $\Delta Y/\Delta O $, and compare them with predictions based
on models of galactic chemical evolution.

\keywords{ISM: abundances, H~{\sc{ii}} regions; galaxies: abundances; galaxies:
  evolution; galaxies: irregular; Galaxy: disk; early universe}
\end{abstract}

\firstsection 
\section{Overview}
The determinations of the helium abundance by mass, $Y$, from metal-poor extragalactic H~{\sc{ii}} regions provide the best method to obtain the primordial helium abundance. During their evolution
galaxies produce a certain amount of helium and oxygen per unit mass that we will call $\Delta Y$
and $\Delta O $. The galaxies less affected by chemical evolution are those that present a large fraction
of their baryonic mass in gaseous form and a small fraction of their baryonic mass in stellar form. These galaxies
when experiencing bursts of star formation present bright metal poor H~{\sc{ii}} regions that have been used to determine their chemical composition.

From a set of $Y$ and $O$ values and assuming a linear relationship it is possible to obtain
$Y_p$ and $\Delta Y/\Delta O $ from the following equation:
\begin{equation}
Y_p  =  Y - { O} \frac{\Delta Y}{\Delta O}
\label{eq:DeltaO}
.\end{equation}

The determinations of $Y_p$ and $\Delta Y/\Delta O $ are important for at least the following reasons: (a) $Y_p$ is one of the pillars of Big Bang cosmology and an accurate determination of $Y_p$ permits to test the Standard Big Bang Nucleosynthesis (SBBN), (b) the models of stellar evolution require an accurate initial $Y$ value; this is given by $Y_p$ plus the additional $\Delta Y$ produced by galactic chemical evolution, which can be estimated based on the observationally determined $\Delta Y/\Delta O $ ratio, (c) the combination of $Y_p$ and $\Delta Y/\Delta O $ is needed to test models of galactic chemical evolution.

Recent reviews on primordial nucleosynthesis have been presented by  Steigman (2007),  Olive (2008), Weinberg (2008), and Pagel (2009). A review on the primordial helium abundance has been presented by Peimbert (2008) and a historical note on the primordial helium abundance has been presented by Peimbert \& Torres-Peimbert (1999).

\section{Recent $Y_p$ determinations}

The best $Y_p$ determinations are those by Izotov, Thuan, \& Stasi\'nska (2007) that amounts to 
$0.2516 \pm 0.0011$ and Peimbert, Luridiana, \& Peimbert (2007a) that amounts to $0.2477 \pm 0.0029$. 
The procedures used by both groups to determine $Y_p$ are very different and it is not easy to make a detailed comparison of all the steps carried out by each of them. We consider that the error presented by Izotov {\it et al.} is a lower limit to the total error because it does not include estimates of some systematic errors. The difference between the central $Y_p$ values derived by both groups is mainly due to the treatment of the temperature structure of the H~{\sc{ii}} regions. Izotov {\it et al.} adopt temperature variations that are smaller than those derived by Peimbert {\it et al.}

To be more specific we can define the temperature structure of the H~{\sc{ii}} regions by means of an average temperature, $T_0$, and a mean square temperature fluctuation, $t^2$ (Peimbert 1967). 
The value of $t^2$ derived by Izotov {\it et al.} (2007) for their sample is about 0.01; while Peimbert {\it et al.} (2007a) obtain a $t^2$ of about 0.026. From the observations adopted by Peimbert {\it et al.} and assuming $t^2$ = 0.000 we obtain $Y_p$ = $0.2523 \pm 0.0027$, and for $t^2$ = $0.01$ we obtain $Y_p = 0.2505$.

The small value of $t^2$ derived by Izotov {\it et al.} (2007) is due to the parameter space used in their Monte Carlo computation where they permitted $T$(He~{\sc{i}}) to vary from 0.95 to 1.0 times the 
$T$(4363/5007) value, which yields a $t^2$ of about 0.01. By allowing their $T$(He~{\sc{ii}}) to vary from 0.80 to 1.0 times the  $T$(4363/5007) value their $t^2$ result would have become higher. This can be seen from their Table 5 where 71 of the 93 spectra corresponded to the lowest $T$(He~{\sc{i}}) allowed by the permitted parameter space of the Monte Carlo computation. A higher $t^2$ value for the Izotov {\it et al.} (2007) sample produces a lower $Y_p$,  reducing the difference with the Peimbert {\it et al.} (2007a) $Y_p$ value.

Two other systematic problems with the Izotov {\it et al.} (2007) determination related with the temperature structure are: (a) that to compute the $N$(O$^{++}$) abundance they adopted the $T$(4363/5007) value which is equivalent to adopt 
$t^2 = 0.00$, and (b) to  compute the once ionized oxygen and nitrogen abundances, $N$(O$^+$) and    
$N$(N$^+$), they adopted the $T$(4363/5007) value, but according to photoionization models and observations of O-poor objects, the temperature in the O$^+$ regions, $T$(O$^+$), is considerably smaller than in the O$^{++}$ regions, typically by about 2000 K for objects with $T$(O$^{++}$) = 16000 K, and reaching 4000 K for the metal poorest H~{\sc{ii}} regions ({\it e. g. } Peimbert, Peimbert, \& Luridiana  2002; Stasi\'nska 1990).

\subsection{Recombination coefficients of the helium I lines}

There are two recent line emissivity estimates to derive the He abundance from recombination lines: one due to
Benjamin, Skillman, \& Smits (1999, 2002), and another by Bauman {\it et al.} (2005) and Porter {\it et al.} (2007, 2009). The difference in the $Y$ values derived from both sets of data amount to about 0.0040, the emissivities by the first group yielding values smaller than those of the second group. According to Porter, Ferland, \& MacAdam (2007), the error introduced in their emissivities by interpolating in temperature the equations provided by them  is smaller than 0.03\%, which translates into an error in $Y_p$ considerably smaller than 0.0001. Moreover according to Porter {\it et al.} (2009) the expected error in their line emissivities amounts to about 0.0010 in the $Y$ derived values. In this review we are adopting the Bauman et al. and Porter et al. emissivities in the presented $Y_p$ values.

\subsection{Beyond case B}

There are at least four processes that modify the level populations of H and He atoms relative to case B and consequently
the  $Y_p$ determination: the optical depth of the He~{\sc{i}} lines, the collisional excitation from the He $2\,^3S$ metastable level, the collisional excitations from the H ground level, and the fluorescent excitation of the  H~{\sc{i}} 
and the He~{\sc{i}} lines, (case D of Luridiana et al. 2009). The last two processes are the least studied of the four.

Luridiana {\it et al.} (2009) have introduced case D into the study of gaseous nebulae. Case D increases slightly the
emissivities of the H~{\sc{i}} and He~{\sc{i}} lines affecting the accuracy of the $Y_p$ determination. There are no published estimates
of the importance of this effect but it depends on the spectra of the ionizing stars, particularly in the region of the 
H~{\sc{i}} and He~{\sc{i}} lines, the spatial distribution of the ionizing stars, the gaseous electron density distribution, the fraction of ionizing photons that escape the nebula, the radial velocity of the gas relative to that of the stars, the nebular turbulence, and the region of the nebula observed. Therefore it requires tailor-made models for each observed object. Peimbert {\it et al.} (2007a) presented a list of thirteen sources of error in the $Y_p$ determination, each of them with systematic and statistical components but dominated by one or the other. The systematic error produced by not having considered Case D should be added to that list.

\section{Comparison of the directly determined $Y_p$ with the $Y_p$ values computed under the assumption of
the SBBN and the observations of $D_p$ and WMAP}

\begin{table}\def~{\hphantom{0}}
  \begin{center}
  \caption{COSMOLOGICAL PREDICTIONS BASED ON SBBN AND OBSERVATIONS 
                                                    FOR $\tau_n$ = 885.7 $\pm$ 0.8 s}
  \label{tab:885.7}
  \begin{tabular}{lcccc}\hline
Method & $Y_p$          &  $D_p$   & $\eta_{10}$  & $\Omega_{b}h^2$  \\
    \hline
$Y_p$   & $0.2477\pm0.0029^a$& $2.93 +2.53-1.06^b$ & $5.625\pm1.81^b$  & $0.02054\pm0.00661^b$    \\
$D_p$   & $0.2479\pm0.0007^b$& $2.82\pm0.28^a$      & $5.764\pm0.360^b$ & $0.02104\pm0.00132^b$   \\
$WMAP$  & $0.2487\pm0.0006^b$& $2.49\pm0.13^b$      & $6.226\pm0.170^b$ & $0.02273\pm0.00062^a$   \\
  \hline 
  \end{tabular}
  
  $^a$Observed value. $^b$Predicted value. {References:} $\tau_n$ Arzumanov {\it et al.} (2000); $Y_p$ Peimbert {\it et al.} (2007a); $D_p$ O'Meara $et al. $(2006); $WMAP$ Dunkley {\it et al.} (2009).
  
  \end{center}
\end{table}

\begin{table}\def~{\hphantom{0}}
  \begin{center}
  \caption{COSMOLOGICAL PREDICTIONS BASED ON SBBN AND OBSERVATIONS 
                                                    FOR $\tau_{n}$ = 878.5 $\pm$ 0.8 s}
  \label{tab:878.5}
  \begin{tabular}{lcccc}\hline
Method & $Y_p$          &  $D_p$   & $\eta_{10}$  & $\Omega_{b}h^2$  \\
    \hline
$Y_p$   & $0.2477\pm0.0029^a$& $2.22 + 1.46-0.71^b$ & $6.688\pm1.81^b$  & $0.02442\pm0.00661^b$   \\
$D_p$   & $0.2462\pm0.0007^b$& $2.82\pm0.28^a$      & $5.764\pm0.360^b$ & $0.02104             \pm0.00132^b$   \\
$WMAP$  & $0.2470\pm0.0006^b$& $2.49\pm0.13^b$      & $6.226\pm0.170^b$ & $0.02273\pm0.00062^a$   \\
  \hline 
  \end{tabular}

   $^a$Observed value. $^b$Predicted value. {References:} same as in Table 1 with the exception of 
$\tau_n$ that comes from Serebrov {\it et al.} (2005, 2008).
  
  \end{center}
\end{table}

To compare the $Y_p$ value with the primordial deuterium abundance $D_p$ (usually expressed as 
$10^5(D/H)_p)$ and with the WMAP results, we will use the framework of the SBBN. The ratio of baryons to photons multiplied by $10^{10}, \eta_{10}$, is given by (Steigman 2006, 2007):
\begin{equation}
\eta_{10} = (273.9 \pm 0.3){\Omega_{b}h^2} 
\label{eq:eta}
,\end{equation}
where $\Omega_{b}$ is the baryon closure parameter, and $h$ is the Hubble parameter. In the range 
$4 < \eta_{10} < 8$ (corresponding to $0.2448 < Y_p < 0.2512)$, $Y_p$ is related to $\eta_{10}$ by (Steigman 2006, 2007):
\begin{equation}
Y_p = 0.2483 \pm 0.0005+ 0.0016(\eta_{10}-6)
\label{eq:helium}
.\end{equation}

In the same $\eta_{10}$ range, the primordial deuterium abundance is given by (Steigman 2006, 2007):
\begin{equation}
10^{5}(D/H)_p = D_p = 46.5 (1 \pm 0.03)(\eta_{10})^{-1.6}
\label{eq:deut} 
.\end{equation}

From the $Y_p$ value by Peimbert {\it et al.} (2007a), the $D_p$ value by O'Meara {\it et al.} (2006), the $\Omega_{b}h^2$ value by Dunkley {\it et al.} (2009), and the previous equations we have produced Table 1. From this table, it follows that within the errors $Y_p$, $D_p$, and the WMAP observations are in very good agreement with the predicted SBBN values.

Equations (2), (3), and (4) were derived under the assumption of a neutron lifetime, $\tau_n$ , of $885.7 \pm 0.8 $ s (Arzumanov {\it et al.} 2000). A recent result by Serebrov {\it et al.} (2005, 2008) yielded a $\tau_n$ of 
$878.5 \pm 0.7 \pm 0.3$ s. This result would lead to a SBBN $Y_p$ value of 0.2470 for the WMAP  $\Omega_{b}h^2$ determined value (Steigman 2007, Mathews {\it et al.} 2005). Mathews {\it et al.} obtain for $\tau_n$ = $881.9 \pm 1.6 $ s, mainly the average of the results by
Arzumanov {\it et al.} (2000) and Serebrov {\it et al.} (2005, 2008), a $Y_p$ value 0.0009 smaller than for $\tau_n$ = $885.7 \pm 0.8 $.

The 9$\sigma$ difference between both $\tau_n$ determinations probably indicates that at least one of them includes
systematic errors that have not yet been sorted out.

From the $Y_p$ by Peimbert {\it et al.} (2007a), the $D_p$ by O'Meara {\it et al.} (2006), and SBBN it is found that for $\tau_n$ = $881.9 \pm 1.6 $ the number of effective neutrino families, $N_{eff}$, is equal to $3.12 \pm 0.23$. Based on the production of the Z particle by electron-positron collisions in the laboratory and taking into account the partial heating of neutrinos produced by electron-positron annihilations during SBBN Mangano {\it et al.} (2002) find that $N_{eff}$ = $3.04$.  The $N_{eff}$ value derived from $Y_p$, $D_p$, and SBBN is in excellent agreement with the value derived by Mangano {\it et al.} (2002).

The $Y_p$ derived from WMAP is based on the very strong assumption of SBBN. It is also possible to derive $Y_p$ from the study of the microwave radiation without assuming SBBN: from the cosmic microwave background radiation ($i. e.$ WMAP + ACBAR + CBI + BOOMERANG), Ichikawa, Sekiguchi and Takahashi (2008a,b) obtain that $Y_p<0.44$, when they also include the information obtained from BAO + SN + HST (baryon acoustic oscillations in the distribution of galaxies, the distance measurements from type Ia supernovae, and the HST value for $H_0$) the constrain improves to $Y_p= 0.25 {+0.10 \atop -0.07}$. Including the expected data from the Planck satellite they predict a reduction on the $Y_p$ error 
of about a factor of four to seven, an error still about four to six times higher than the one estimated from the best $Y_p$ determinations based on metal poor H~{\sc{ii}} region observations. 

\section{The $\Delta Y/\Delta O $ ratio}

To determine the $Y_p$  value from a set of metal poor H~{\sc{ii}}  regions it is necessary to estimate the fraction of helium, present in the interstellar medium of the galaxy where each H~{\sc{ii}} region is located, produced by galactic chemical evolution. From observations of metal poor extragalactic H~{\sc{ii}}  regions it has been found that the $Y$
versus $O$ observations  can be fitted with a straight line given by $\Delta Y/\Delta O $ and equation (1) 
has been used often to derive $Y_p$. A straight line is predicted by chemical evolution models of metal poor galaxies with the same initial mass function, a given set of stellar yields, and different star formation rates. To obtain different
$\Delta Y/\Delta O $  from the models it is necessary to change the initial mass function (for example the maximum mass allowed or the slope at high masses), or the adopted yields.

The observational $Y_p$,  $\Delta Y$, and $\Delta O $ are affected by different amounts in the presence of temperature
variations, while $Y_p$ diminishes by 0.0046 due to temperature variations in the sample of Peimbert et al. (2007a),
$\Delta Y$ is slightly affected and $\Delta O $ is strongly affected by temperature variations ({\it e. g. } Carigi \& Peimbert 2008, Table 3). In addition a fraction of $O$ is embedded in dust grains, fraction that needs to be estimated to compare with models of galactic chemical evolution.

The importance of $\Delta Y/\Delta O $ is two fold: it permits us to obtain a more accurate $Y_p$ value, and
permits us to test for the presence of large temperature variations in gaseous nebulae when comparing nebular values with stellar ones.

\subsection{The gaseous O/H determination}

There are two methods to derive the gaseous O/H ratio, from the $I(4363)/I(5007)$ ratio together with  the
$I(3727)/I({\rm H}\beta)$ and the $I(5007)/I({\rm H}\beta)$ line ratios, the so called 
$T(4363)$ method, and from the intensity ratio of O~{\sc{ii}} 
recombination lines to  H~{\sc{i}}  recombination lines that has been called the O~{\sc{ii}}$_{RL}$ 
method by Peimbert {\it et al.} (2007b). The O~{\sc{ii}}$_{RL}$ method usually provides higher O/H ratios by 0.15 to about
0.3 dex, this difference is due to temperature variations inside the observed volume and is smaller for metal poor 
H~{\sc{ii}} regions and higher for metal rich H~{\sc{ii}} regions. The $T(4363)$ method in the presence of temperature 
variations produces a systematic effect that lowers the O/H abundances relative to the real ones. On the other hand the
O~{\sc{ii}}$_{RL}$ method is independent of the temperature structure.

\subsection{The total O/H determination}

Another factor that has to be taken into account to obtain the total O/H ratio is the fraction of
O atoms trapped in dust grains. Esteban {\it et al.} (1998), based on the depletion of Fe, Mg and Si in the Orion
nebula, estimated that the fraction of oxygen atoms trapped in dust grains amounts to 0.08 dex. 
Mesa-Delgado {\it et al.} (2009) have estimated that the fraction of O atoms trapped in dust grains in the Orion nebula amounts to 0.12 $\pm$ 0.03 dex; this result is based on three different methods: a) by comparing the O abundances of the B stars of the Orion association with the O abundance of the Orion nebula, b) from the depletion of Fe, Mg and Si in the nebula and assuming that these atoms are combined with molecules containing O, and c) by comparing the
shock and nebular abundances of the material associated with the Herbig-Haro object 202.

Rodr\'{\i}guez \& Rubin (2005) have estimated the fraction of Fe atoms in galactic and extragalactic H~{\sc{ii}} regions in the gaseous phase, the depletions derived for the different objects define a trend of increasing depletion at higher metallicities; while the Galactic H~{\sc{ii}} regions show less than than 5\% of their Fe atoms in the gas phase, the extragalactic ones (LMC 30 Doradus, SMC N88A, and SBS 0335-052) have somewhat lower depletions. Izotov {\it et al.} (2006) find a slight increase of Ne/O with increasing metallicity, which they interpret as due to a moderate depletion of O onto grains in the most metal-rich galaxies, they conclude that this O/Ne depletion corresponds to $\sim$ 20\% of oxygen locked in the dust grains in the highest-metallicity H~{\sc{ii}}  regions of their sample, while no significant depletion would be present in the H~{\sc{ii}} regions with lower metallicity.  Peimbert \& Peimbert (2010) based on the Fe/O abundances of Galactic and extragalactic H~{\sc{ii}} regions estimate that for objects in the 
$8.3 < 12$ + log O/H $< 8.9$ range the fraction of O atoms trapped by dust grains amounts to $0.12 \pm 0.03$ dex, for objects in the $7.7 < 12$ + log O/H $< 8.3$ range amounts to $0.09 \pm 0.03$ dex, and for objects in the 
$7.3 < 12$ + log O/H $< 7.7$ range amounts to $0.06 \pm 0.03$ dex.

\subsection{Extrapolation of the $Y$ determinations to the value of $Y_p$, or the $O$ ($\Delta Y/\Delta O $) correction}

From chemical evolution  models of different galaxies it is found 
that $\Delta Y/\Delta O $ depends on the initial mass function (IMF), the star formation rate, the age, and the $O$ value of the galaxy in question. Peimbert {\it et al.} (2007b) have found that $\Delta Y/\Delta O $ is well fitted by a constant value for objects with the same IMF, the same age, and an O abundance smaller than $4 \times 10^{-3}$. This result is consistent with the custom of using a constant value for  $\Delta Y/\Delta O $ to fit observational data.

To obtain an accurate $Y_p$ value, a reliable determination of $\Delta Y/\Delta O $ for O-poor objects is needed. The $\Delta Y/\Delta O $ value derived by Peimbert, Peimbert, \& Ruiz (2000) from observational results and models of chemical evolution of galaxies amounts to $ 3.5 \pm 0.9$. More recent results are those by Peimbert (2003) who finds $2.93 \pm 0.85$ from observations of 30 Dor and NGC 346, and by Izotov \& Thuan (2004) who, from the observations of 82 H~{\sc{ii}} regions, find $\Delta Y/\Delta O $ = $4.3 \pm 0.7$. Peimbert {\it et al.}  (2007a) have recomputed this value by taking into account two systematic effects not considered by  Izotov \& Thuan: the fraction of oxygen trapped in dust grains, and the increase in the O abundances due to the presence of temperature variations. From these considerations they obtained for the Izotov \& Thuan sample that $\Delta Y/\Delta O $ =  $3.2  \pm  0.7$.

On the other hand Peimbert {\it et al.} (2007b) from chemical evolution models with different histories of galactic inflows and outflows for objects with $O < 4 \times 10^{-3}$ find that $2.4 < \Delta Y/\Delta O  < 4.0$. From the theoretical and observational results Peimbert et al. (2007a) adopted a value of $\Delta Y/\Delta O $ =  $3.3 \pm 0.7$, that they used with the $Y$ and $O$ determinations from each object to obtain the $Y_p$ value. 

\subsection{Comparison of the oxygen abundances of the ISM of the solar vicinity with those of the Sun
and F and G stars of the solar vicinity}

In addition to the evidence presented in section 2 in favor
of large $t^2$ values, and consequently in favor of the O~{\sc{ii}}$_{RL}$ 
method, there is another independent test that can be used to discriminate 
between the $T$(4363) method and the O~{\sc{ii}}$_{RL}$ method that consists
in the comparison of stellar and  H~{\sc{ii}} region abundances of the 
solar vicinity. 

Esteban {\it et al.} (2005) determined that the gaseous O/H value derived from  H~{\sc{ii}} regions of the solar vicinity 
amounts to 12 + log (O/H) = 8.69, and including the fraction of O atoms tied up in dust grains it is
obtained that 12 + log (O/H) = 8.81 $\pm~0.04$ for the O/H value of the ISM of the solar vicinity. 
Alternatively from the 
protosolar value by Asplund {\it et al.} (2009), 
that amounts to 12 + log(O/H) = 8.71, and taking into account
the increase of the O/H ratio due to galactic chemical evolution since
the Sun was formed, that according to the chemical
evolution model of the Galaxy by Carigi {\it et al.} (2005) amounts to 0.13~dex, 
we obtain an O/H value of 8.84 $\pm~0.04$~dex, in excellent agreement 
with the value based on the O~{\sc{ii}}$_{RL}$ method. In this comparison 
we are assuming that the solar abundances are representative of the abundances 
of the solar vicinity ISM when the Sun was formed.

There are two other estimates of the 
O/H value in the ISM that can be made from 
observations of F and G stars of the solar vicinity. According to 
Allende-Prieto {\it et al.} (2004) the Sun appears deficient by roughly 
0.1 dex in O, Si, Ca, Sc, Ti, Y, Ce, Nd, and Eu, compared with its 
immediate neighbors with similar iron abundances; by assuming that the O abundances of the solar
immediate neighbors are more representative of the local ISM than the solar one, and by adding this 
0.1 dex difference to the 
protosolar solar value by Asplund {\it et al.} (2009) we obtain that the present value of the ISM has to be higher than
12 + log O/H = 8.81.
A similar result is obtained from the data
by Bensby \& Feltzing (2005) who obtain for the six most O-rich 
thin-disk F and G dwarfs of the solar vicinity an average  
[O/H] = 0.16; by assuming their value as representative of the 
present day ISM of the solar vicinity we find 12 + log O/H = 8.87. 
Both results are in good agreement with the O/H value derived 
from the O~{\sc{ii}}$_{RL}$ method.

\subsection{Comparison of $\Delta Y/\Delta Z $ values of the Galactic H~{\sc{ii}} region M17 with K dwarfs
of the solar vicinity}

The best Galactic H~{\sc{ii}} region to determine the He/H ratio is M17 because it
contains a very small fraction of neutral helium and the error introduced by correcting
for its presence is very small. Carigi \& Peimbert (2008) obtained from observations for M17 (and adopting a $Y_p$ value) a value of $\Delta Y/\Delta Z = 1.97 \pm 0.41$ for $t^2 = 0.036 \pm 0.013$, where $Y$ and $Z$ are the helium and
heavy elements by unit mass; by correcting this value considering that the fraction of O trapped in dust amounts to 0.12 dex instead of 0.08 dex (Mesa-Delgado {\it et al.} 2009) we obtain $\Delta Y/\Delta Z~=~1.77~\pm~0.37$. This $\Delta Y/\Delta Z$ value is in agreement with two independent $\Delta Y/\Delta Z$ 
determinations derived from K dwarf stars of the solar vicinity that amount to $2.1~\pm~0.4$ (Jim\'enez {\it et al.} 2003)~and to $2.1~\pm~0.9$ (Casagrande {\it et al.} 2007). 

On the other hand 
the value $\Delta Y/\Delta Z ~=~3.6 ~\pm~0.68$ derived from collisionally excited lines of M17 under
the assumption of $t^2 = 0.00$ by Carigi and Peimbert (2008) (value corrected for a fraction of O trapped by dust grains of
0.12 dex) is not in agreement with the values derived from K dwarf stars of the solar neighborhood. 

\begin{figure}[b]
\begin{center}
 \includegraphics[width=2.4in]{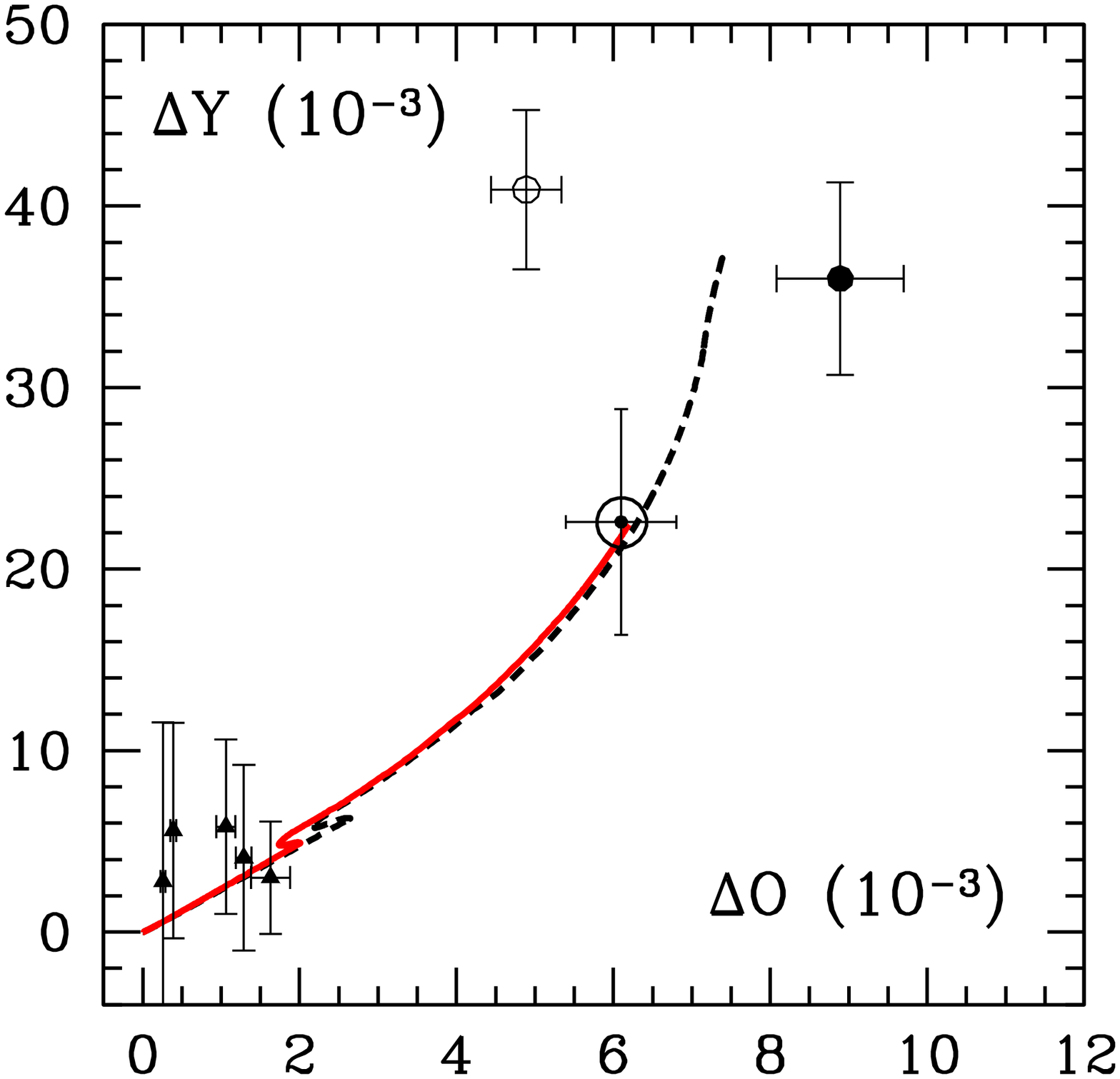}  
\includegraphics[width=2.4in]{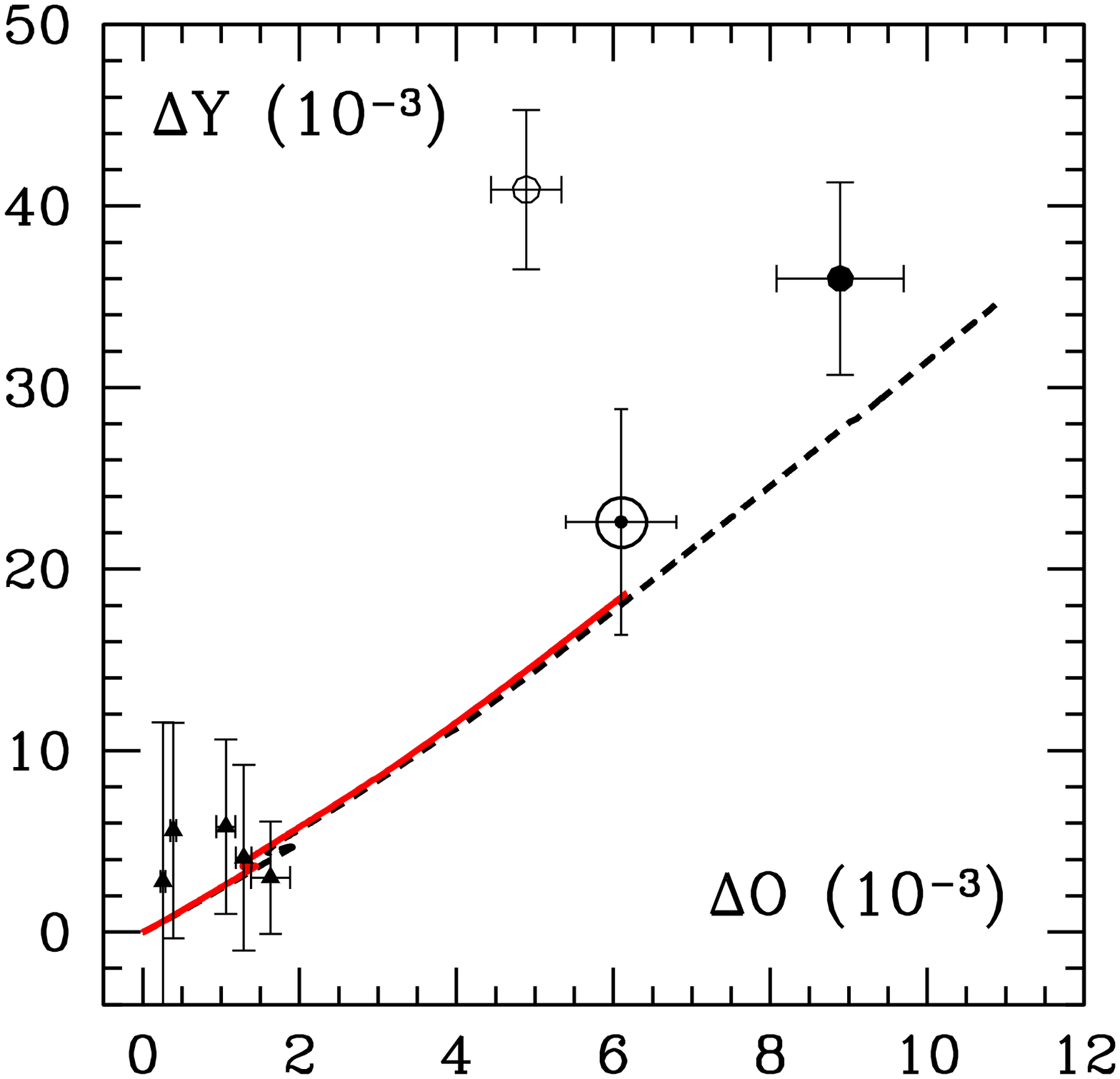} 
 \caption{In the two panels the origin corresponds to $Y_p$ = 0.2477, the triangles represent the $Y$ and $O$ values derived by Peimbert {\it et al.} (2007a) for five metal poor extragalactic H~{\sc{ii}} regions, the open and filled circles represent the M17 values for $t^2 = 0.00$, and $t^2 = 0.036$ respectively from Carigi \& Peimbert (2008), the large open circle with a dot in the center corresponds to the
presolar values by Asplund {\it et al.} (2009). The left panel presents two chemical evolution models by Carigi \& Peimbert (2010), the solid line is a model for the Galactic disk with a time span from the formation of the Galaxy to the formation of the Sun at a galactocentric distance of 8 kpc, while the dashed line corresponds to a model for the Galactic disk with a time span from the formation of the Galaxy to the present at a galactocentric distance of 6.75 kpc, the distance of M17 to the galactic center; the stellar yields for the two panels are different, see subsection 4.6 for further details.}
   \label{fig1}
\end{center}
\end{figure}

\subsection{Comparison of $Y_p$ and $\Delta Y/\Delta O $ with models of chemical evolution for the disk of the Galaxy}

To compare Galactic chemical evolution models of $Y$ and $O$ with observations we need to use the best determinations
available of the abundances of these elements. We consider that the two most accurate Galactic $Y$ and $O$ determinations are the presolar values (Asplund {\it et al.} 2009) and the M17 H~{\sc{ii}} region values (Carigi \& Peimbert 2008).

Carigi {\it et al.} (2005) presented chemical evolution models for the disk of the Galaxy that fit the slope and the absolute value of the O/H gradient, these models are also successful in reproducing the C/O gradient derived from
H~{\sc{ii}} regions and the C/O versus O/H evolution history of the solar vicinity obtained from stellar observations.  In Figure 1 we present chemical evolution models by Carigi \& Peimbert (2010) for the $Y$ and $O$ abundances of the Galactic disk for two sets of stellar yields where they have added the presolar values by Asplund et al. (2009) for comparison. These models are based on the same assumptions than those adopted by Carigi and Peimbert (2008). The only difference between the models in the left panel and those in the right one, is that for massive stars with $Z > 0.004$ the ones in the left use the yields by Maeder (1992) with high mass loss, while the ones on the right use the yields by Hirschi, Meynet, \& Maeder (2005) with low mass loss. There are at least three conclusions that we can extract from the figure: (a) the fit for the presolar values is very good for the two sets of yields, (b) the fit for M17 for $t^2 = 0.036 \pm 0.013$
is good, but it can be improved by assuming a set of yields intermediate between the two sets used; a similar result in favor of an intermediate set of yields was obtained by Cescutti {\it et al.} (2009) based on C/O observations for stars in the bulge of the Galaxy, (c) the models do not fit M17 for $t^2 = 0.00$, this result is in agreement with those by Esteban {\it et al.} (2005, 2009) and Peimbert {\it et al.} (2007a) who find large $t^2$ values for Galactic and extragalactic H~{\sc{ii}} regions. Moreover the $O$ and $Y$ abundances for the presolar material and for M17 are derived from independent methods, and they are fitted by the same chemical evolution model. This fit provides us with a consistency check on the gaseous nebulae helium and oxygen abundance determinations based on large $t^2$ values.

\section{Conclusions}
 
During the last 50 years the determination of $Y_p$ has been very important for the study of cosmology, stellar evolution, and the chemical evolution of galaxies. To determine $Y_p$ it is necessary to determine accurate atomic parameters and the physical conditions inside ionized gaseous nebulae.

During the last five decades the accuracy of the $Y_p$ determination has increased considerably, and during the last two decades the differences among the best $Y_p$ determinations have been due to systematic effects ({\it e. g. } Olive \& Skillman 2004, Peimbert 2008). These systematic effects have been gradually understood, particularly during the last few years.

The best $Y_p$ determination available, that by Peimbert {\it et al.} (2007a), is in agreement with the $D_p$ determination and with the WMAP observations under the assumption of SBBN. The errors in the $Y_p$ determination are still large and there is room for non-standard physics.

Similarly to improve the accuracy of the $Y_p$ value derived under the assumption of SBBN and the $\Omega_{b}h^2$ derived from the background radiation, a new determination of the neutron lifetime is needed to sort out the difference between the $\tau_n$  obtained by Arzumanov {\it et al.} (2000) and the $\tau_n$ obtained by Serebrov {\it et al.} (2005, 2008).

To improve the accuracy of the $Y_p$ determination based on metal poor H~{\sc{ii}} regions, the following steps should be taken in the near future: (a) to obtain new observations of high spectral resolution of metal poor H~{\sc{ii}}  regions, those with $0.0005 < Z < 0.001$ to reduce the effect of the collisional excitation of the Balmer lines, which for the present day $Y_p$ determinations is one of the two main sources of error; (b) to determine the temperature of the H~{\sc{ii}} regions based on a large number of He~{\sc{i}} lines observed with high accuracy, and to try to avoid the use of 
$T$(4363/5007) temperatures that weigh preferentially the regions of higher temperature than the average one, effect that artificially increases the $Y_p$ determinations; (c) additional efforts should be made to understand the mechanisms that produce temperature variations in giant H~{\sc{ii}} regions and once they are understood they should be incorporated into photoionization models; (d) the He~{\sc{i}} recombination coefficients should be computed again with an accuracy higher than that of the last two determinations; (e) the computation of tailor-made photoionization models for extragalactic  H~{\sc{ii}} regions including the effect of case D on the intensity of the H~{\sc{i}} and He~{\sc{i}} lines should be carried out. 

The observed $\Delta Y/\Delta O $ ratios provide us with strong constrains for models of galactic chemical evolution.
The gaseous O abundances have to be corrected by the fraction of oxygen embedded in dust grains and by the effect of temperature variations, these two corrections together amount to about 0.2 to 0.4 dex. Chemical evolution models for the Galactic disk are able to reproduce the observed $Y$ and $O$ presolar values and the $Y$ and $O$  values derived for M17
based on H, He and O recombination lines, but not the M17 $Y$ and $O$ values derived from $T$(4363/5007) and O collisionally excited lines under the assumption of $t^2 = 0.00$. This result provides a consistency check in favor of the presence of large temperature variations in H~{\sc{ii}} regions and in favor of the 
$Y_p$ determinations based on the $T$(He~{\sc{i}}) and $t^2$ values derived from He~{\sc{i}} recombination lines.

\begin{acknowledgments}
We wish to thank Gary Ferland, Evan Skillman and Gary Steigman for several fruitful discussions. We are grateful to the SOC for an outstanding meeting and to the LOC for their warm hospitality. We would also like to
acknowledge partial support received from CONACyT grant 46904.
\end{acknowledgments}

\end{document}